\documentstyle[aps]{revtex}

\begin{document}
\author{E. J. FERRER$^{\thanks{{\it On leave from Department of Physics,
SUNY-Fredonia, NY 14063, USA}}}$}
\address{Institute for Space Studies of Catalonia, CSIC, Edif. Nexus 201, Gran Capita%
\\
2-4, 08034 Barcelona, Spain}
\author{and}
\author{V. de la INCERA$^{*}$}
\address{University of Barcelona, Department of Structure and Constituents of Matter,%
\\
Diagonal 647, 08028 Barcelona, Spain}
\title{{\bf On Magnetic Catalysis and Gauge Symmetry Breaking}}
\maketitle

\begin{abstract}
Non-perturbative effects of constant magnetic fields in a Higgs-Yukawa gauge
model are studied using the extremum equations of the effective action for
composite operators. It is found that the magnetic field induces a Higgs
condensate, a fermion-antifermion condensate, and a fermion dynamical mass,
hence breaking the discrete chiral symmetry of the theory. The results imply
that for a non-simple group extension of the present model, the external
magnetic field would either induce or reinforce gauge symmetry breaking.
Possible cosmological applications of these results in the electroweak phase
transition are suggested.
\end{abstract}

Symmetry behavior in quantum field theories under the influence of external
fields has long been a topic of intensive study in theoretical physics\cite
{books}. In the present paper we are interested in particular in
non-perturbative effects produced by external magnetic fields in gauge
theories with scalars. Our main claim is that for theories with non-simple
gauge group, scalar-scalar and scalar-fermion interactions, the magnetic
field reinforces gauge symmetry breaking.

The observation of large-scale galactic magnetic fields in a number of
galaxies, in galactic halos, and in clusters of galaxies \cite{observations}
has recently stimulated a large number of works trying to explain the
physical mechanism responsible for the origin of these fields. Many of the
proposed generating mechanisms have compelling arguments in favor of the
existence of strong primordial magnetic fields (for a review of cosmological
generating mechanisms see \cite{enq} and references therein). Since
primordial magnetic fields could play a significant role in particle
cosmology, the investigations on the theme have recently boomed. In this
context, the implications of a magnetic-field-driven gauge symmetry breaking
mechanism may be important.

Several years before this renewed interest in cosmological magnetic fields,
Amb\o jrn and Olesen \cite{amb-olesen} considered the electroweak model in
the presence of a constant magnetic field. Assuming certain special values
of the couplings, they obtained a W- and Z-condensate solution forming a
lattice of abelian vortex lines for a range of magnetic fields lying between 
$\frac{m_{W}^{2}}{e}$ and $\frac{m_{W}^{2}\cos ^{2}\theta }{e}.$ At even
larger values of the magnetic fields they found that the phase transition to
a symmetric phase can be reached at temperatures lower than the critical one
at zero field. This result realizes, although due to a totally different
reason, an old suggestion \cite{salam} that large magnetic fields could
induce the transition from the broken to the unbroken phase in the
electroweak system.

More recently, the ground state of the electroweak theory in the presence of
a hypermagnetic field has been investigated using either numerical or
perturbative calculations \cite{Giovannini-Shaposhnikov}-\cite{Skalazub}.
The main motivation of these papers was to study the possibility that a
hypermagnetic field could allow the realization of baryogenesis within the
Standard Model \cite{Giovannini-Shaposhnikov}. Even though the original
results \cite{Giovannini-Shaposhnikov} for the upper bound of the Higgs mass
needed to have baryogenesis in the SM were quite optimist, it was quickly
realized that higher loop effects \cite{hyperB},\cite{Skalazub} and
numerical non-perturbative calculations \cite{hyperb1} would significantly
weaken the transition. Moreover, posterior studies on which certain
subtleties of the theory- like the magnetic dipole moment of the sphaleron 
\cite{comelli} or ring diagrams contributions to the high-temperature
effective potential \cite{Skalazub}- were taken into account, concluded that
albeit the hypermagnetic field strengthen the first order character of the
phase transition, it is not enough to satisfy the SM baryogenesis condition 
\cite{shap}

\begin{equation}
\frac{v\left( T_{c}\right) }{T_{c}}\geq 1,  \label{1}
\end{equation}
with $v\left( T_{c}\right) $ the Higgs vacuum expectation value (vev) at the
critical temperature $T_{c}$ of the electroweak phase transition.

When a non-perturbative analytic approach is used to study field theories in
external magnetic fields, new non-trivial effects are found. An important
example of these non-perturbative effects is the formation of a chiral
symmetry breaking fermion condensate $<\overline{\psi }\psi >$ and of a
dynamically generated fermion mass in the presence of an external magnetic
field, known in the literature as magnetic catalysis\cite{mirans-gus-shoko}.
This phenomenon, which has proven to be rather universal and model
independent, has recently attracted a lot of attention\cite{ackley}-\cite
{composite}.

On normal circumstances massless fermions can condensate and acquire a
dynamical mass, but the condensate appears only for sufficiently strong
coupling between fermions. The new feature when a magnetic field is present
is that it favors (catalyzes) the symmetry breaking by reducing to the
weakest attractive coupling the strength of the interaction needed to break
the symmetry. The essence of this effect is that the fermions in the lowest
Landau level (LLL) constitute the effective fermionic degrees of freedom
whose dynamics dominates the long wavelength behavior of the system. The
phenomenon is driven by the fact that massless fermions acquire an energy
gap in the presence of a magnetic field, but there is no energy gap between
the vacuum and the LLL fermions. Then, in the infrared region, the dynamics
of the LLL fermions dominates the fermion propagator, making it essentially
D-2 dimensional. This effective dimensional reduction strengthen the fermion
pairing dynamics\cite{mirans-gus-shoko},\cite{simon} giving rise to a
fermion condensate.

It is worth to mention that the phenomenon of magnetic catalysis is not only
interesting from a purely fundamental point of view, but it has potential
application in condensed matter \cite{mavro}-\cite{vv3} and cosmology \cite
{vv1}. For instance, it has been recently speculated that the generation of
mass through magnetic catalysis in lower dimensional models \cite{mav},\cite
{cond-mat}, or in four-dimensional models with boundaries \cite{vv3}, could
be behind the physical mechanism explaining the observed scaling of the
thermal conductivity in superconducting cuprates with an externally applied
magnetic field \cite{krishana}. On the other hand, the magnetic catalysis
could influence the character of the electroweak phase transition as
suggested by the results of ref. \cite{vv1}.

In the present paper we consider a simple model field theory with the aim of
investigating in a self-consistent way how scalar-scalar and fermion-scalar
interactions in the presence of an external magnetic field can influence the
stability of the vacuum. It is not intended as a realistic theory, but
rather as an example of a large class of theories with scalar fields, on
which dynamical symmetry breaking (either chiral or gauge) can be catalyzed
by an external magnetic field. In this sense, it could be useful for
condensed matter, as well as for cosmological applications. If this toy
model is extended to include a non-simple gauge group theory, as for
instance the electroweak model, the results of this paper could provide a
scenario on which, in contrast to the effect found by Amb\o jrn and Olesen 
\cite{amb-olesen}, an external magnetic field could either induce gauge
symmetry breaking or reinforce it, in the case it already exists, through
non-perturbative effects.

Let us consider the following theory of gauge, fermionic and real scalar
fields described by the Higgs-Yukawa Lagrangian density

\begin{equation}
L=-\frac{1}{4}F^{\mu \nu }F_{\mu \nu }+i\overline{\psi }\gamma ^{\mu
}\partial _{\mu }\psi +g\overline{\psi }\gamma ^{\mu }\psi A_{\mu }-\frac{1}{%
2}\partial _{\mu }\varphi \partial ^{\mu }\varphi -\frac{\lambda }{4!}%
\varphi ^{4}-\frac{\mu ^{2}}{2}\varphi ^{2}-\lambda _{y}\varphi \overline{%
\psi }\psi  \label{e1}
\end{equation}
This theory has a U(1) gauge symmetry, 
\begin{eqnarray}
A_{\mu } &\rightarrow &A_{\mu }+\frac{1}{g}\partial _{\mu }\alpha (x) 
\nonumber \\
\psi &\rightarrow &e^{i\alpha (x)}\psi ,  \label{ee1}
\end{eqnarray}
a fermion number global symmetry 
\begin{equation}
\psi \rightarrow e^{i\theta }\psi ,  \label{ee2}
\end{equation}
and a discrete chiral symmetry 
\begin{equation}
\psi \rightarrow \gamma _{_{5}}\psi ,\qquad \overline{\psi }\rightarrow -%
\overline{\psi }\gamma _{_{5}},\qquad \varphi \rightarrow -\varphi
\label{e2}
\end{equation}

Note that a fermion mass term $m\overline{\psi }\psi $ is forbidden, since
it is invariant under (\ref{ee1}) and (\ref{ee2}), but not under the
discrete chiral symmetry (\ref{e2}).

To study the vacuum solutions that could arise in the theory (\ref{e1})
under the influence of an external constant magnetic field $B$, we need to
solve the extremum equations of the effective action $\Gamma $ for composite
operators\cite{jackiw},\cite{miransbook}

\begin{eqnarray}
\frac{\delta \Gamma (\varphi _{c},\overline{G})}{\delta \overline{G}} &=&0,\;
\label{e3} \\
\frac{\delta \Gamma (\varphi _{c},\overline{G})}{\delta \varphi _{c}} &=&0
\label{e4}
\end{eqnarray}
where $\overline{G}(x,x)=\sigma (x)=\left\langle 0\mid \overline{\psi }%
(x)\psi (x)\mid 0\right\rangle $ is a composite fermion-antifermion field,
and $\varphi _{c}$ represents the vev of the Higgs field. Thanks to the
discrete chiral symmetry (\ref{e2}), it is enough to consider only one
composite field. We choose the composite field $\overline{G}(x,x),$ ignoring
the second possible one, $\pi (x)=\left\langle 0\mid \overline{\psi }%
(x)i\gamma _{5}\psi (x)\mid 0\right\rangle ,$ since the effective action can
be a function only of the chirally invariant combination $\rho ^{2}=\sigma
^{2}+\pi ^{2}$.

The loop expansion of the effective action $\Gamma $ for composite operators 
\cite{jackiw},\cite{miransbook} can be expressed as

\begin{equation}
\Gamma \left( \overline{G},\varphi _{c}\right) =S\left( \varphi _{c}\right)
-iTr\ln \overline{G}^{-1}+i\frac{1}{2}Tr\ln D^{-1}+i\frac{1}{2}Tr\ln \Delta
^{-1}-iTr\left[ G^{-1}\left( \varphi _{c}\right) \overline{G}\right] +\Gamma
_{2}\left( \overline{G},\varphi _{c}\right) +C  \label{e5}
\end{equation}
In Eq. (\ref{e5}) $C$ is a constant and $S\left( \varphi _{c}\right) $ is
the classical action evaluated in the scalar vev (Higgs condensate) $\varphi
_{c}.$ The bar on the fermion propagator $\overline{G}\left( x,y\right) $
means that it is taken full, while the non-bar notation indicates free
propagators, as it is the case for the gauge propagator $D_{\mu \nu
}(x-y)=\int \frac{d^{4}q}{(2\pi )^{4}}\frac{e^{iq\cdot (x-x^{\prime })}}{%
q^{2}-i\epsilon }\left( g_{\mu \nu }-(1-\xi )\frac{q_{\mu }q_{\nu }}{%
q^{2}-i\epsilon }\right) ,$ and the scalar one $\Delta (x-y)=\int \frac{%
d^{4}q}{(2\pi )^{4}}\frac{e^{iq\cdot (x-x^{\prime })}}{q^{2}+M^{2}-i\epsilon 
}$, with $M^{2}=\frac{\lambda }{2}\varphi _{c}^{2}+\mu ^{2}.$ In general $%
\Gamma _{2}\left( \overline{G},\varphi _{c}\right) $ represents the sum of
two and higher loop two-particle irreducible vacuum diagrams. In the current
approximation, as all propagators but the fermion's are taken free, $\Gamma
_{2}$ is two-particle irreducible with respect to fermion lines only \cite
{miransbook}. In the present weakly coupled theory one can use the lowest
(two-loop) approximation for $\Gamma _{2}$. This corresponds to the so
called quenched ladder approximation, on which all vertices are taken bare.
In this case $\Gamma _{2}$ is

\begin{eqnarray}
\Gamma _{2}\left( \overline{G},\varphi _{c}\right) &=&\frac{g^{2}}{2}\int
d^{4}xd^{4}ytr\left[ \overline{G}\left( x,y\right) \gamma ^{\mu }\overline{G}%
\left( y,x\right) \gamma ^{\nu }D_{\mu \nu }(x,y)\right]  \nonumber \\
&&-\frac{g^{2}}{2}\int d^{4}xd^{4}ytr\left( \gamma ^{\mu }\overline{G}\left(
x,x\right) \right) D_{\mu \nu }(x-y)tr\left( \gamma ^{\nu }\overline{G}%
\left( y,y\right) \right)  \nonumber \\
&&+\frac{\lambda _{y}^{2}}{2}\int d^{4}xd^{4}ytr\left[ \overline{G}\left(
x,y\right) \overline{G}\left( y,x\right) \Delta (x,y)\right]  \nonumber \\
&&-\frac{\lambda _{y}^{2}}{2}\int d^{4}xd^{4}ytr\left( \overline{G}\left(
x,x\right) \right) \Delta (x-y)tr\left( \overline{G}\left( y,y\right) \right)
\label{e6}
\end{eqnarray}

The extremum equations $\left( \ref{e3}\right) $ and $\left( \ref{e4}\right) 
$ correspond, respectively, to the Schwinger-Dyson (SD) equation for the
fermion self-energy operator $\Sigma $ (gap equation)$,\ $and to the usual
minimum equation for the expectation value of the scalar field, which in the
presence of the magnetic field has to be determined in a self-consistent
way, that is, simultaneously with the gap equation.

Although we have introduced a bare scalar mass $\mu $ in (\ref{e1}), because
we are interested in the possibility of a dynamically generated scalar mass,
we take the limit $\mu \rightarrow 0$ at the end of our calculations.

The second to fifth terms in the effective action $\left( \ref{e5}\right) $
correspond to one-loop contribution. Their evaluation is quite
straightforward (the scalar self-interaction can be renormalized in the
usual way \cite{jackiw-74}), with the exception of the fermion
contributions, which contain the background magnetic field. Then, let us
calculate explicitly the one-loop fermion contribution coming from the term 
\begin{equation}
\Gamma _{f}^{(1)}=-iTr\left[ G^{-1}\left( \varphi _{c}\right) \overline{G}%
\right]   \label{e7}
\end{equation}
in (\ref{e5}). Here $\overline{G}\left( x,y\right) $ denotes the full
fermion propagator, which can be written as \cite{ackley}

\begin{equation}
\overline{G}\left( x,y\right) =\sum\limits_{k}\int \frac{dp_{0}dp_{2}dp_{3}}{%
\left( 2\pi \right) ^{4}}E_{p}\left( x\right) \left( \frac{1}{\gamma .%
\overline{p}+\Sigma (p)}\right) \overline{E}_{p}\left( y\right)  \label{e9}
\end{equation}
and $G^{-1}\left( \varphi _{c}\right) $ is the free fermion inverse
propagator in the presence of a constant magnetic field $B$ along the third
axis$,$

\begin{equation}
G^{-1}\left( x,y,\varphi _{c}\right) =\sum\limits_{k}\int \frac{%
dp_{0}dp_{2}dp_{3}}{\left( 2\pi \right) ^{4}}E_{p}\left( x\right) \left(
\gamma .\overline{p}+m_{0}\right) \overline{E}_{p}\left( y\right) ,
\label{e8}
\end{equation}
with $\overline{p}=(p_{0},0,-\sqrt{2gBk},p_{3}),$ and $m_{0}=\lambda
_{y}\varphi _{c}$, the fermion mass appearing after the shift $\varphi
\rightarrow \varphi +\varphi _{c}$ in the Higgs field.

In the above equations we have introduced the Ritus' $E_{p}$ functions \cite
{ritus}. These orthonormal function-matrices provide an alternative method
to the Schwinger's approach to problems of QFT on electromagnetic backgrounds%
\footnote{%
For an application of Ritus' method to the QED Schwinger-Dyson equation in a
magnetic field see ref.\cite{ackley}.}.

The $E_{p}$ representation is obtained forming the eigenfunction-matrices of
the fermion mass operator

\begin{equation}
E_{p}(x)=\sum\limits_{\sigma }E_{p\sigma }(x)\Delta (\sigma ),  \label{e14}
\end{equation}
where 
\begin{equation}
\Delta (\sigma )=diag(\delta _{\sigma 1},\delta _{\sigma -1},\delta _{\sigma
1},\delta _{\sigma -1}),\qquad \sigma =\pm 1,  \label{e15}
\end{equation}
and the $E_{p\sigma }$ functions are given by

\begin{equation}
E_{p\sigma }(x)=N(n)e^{i(p_{0}x^{0}+p_{2}x^{2}+p_{3}x^{3})}D_{n}(\rho )
\label{e11}
\end{equation}
with $D_{n}(\rho )$ being the parabolic cylinder functions\cite{handbook} of
argument $\rho =\sqrt{2gB}(x_{1}-\frac{p_{2}}{gB})$ and positive integer
index

\begin{equation}
n=n(k,\sigma )\equiv k+\frac{\sigma }{2}-\frac{1}{2}\;\quad n=0,1,2,...,
\label{e13}
\end{equation}
and $N(n)=(4\pi gB)^{\frac{1}{4}}/\sqrt{n!}$ being a normalization factor.
Here $p$ represents the set $(p_{0},p_{2,}p_{3},k),$ which determines the
eigenvalue $\overline{p}^{2}=-p_{0}^{2}+p_{3}^{2}+2gBk$ in $(\gamma ^{\mu
}\left( i\partial _{\mu }-gA_{\mu }\right) )^{2}\psi _{p}=\overline{p}%
^{2}\psi _{p}$ (for details and notation see \cite{ackley} and \cite{vv1})$.$
In Eq. $\left( \ref{e11}\right) $ we are considering the case of a purely
magnetic field background (crossed field case) directed along the
z-direction (without loss of generality we assume that $sign(gB)=1)$.

One can easily check that the $E_{p}$ functions are orthonormal

\begin{equation}
\int d^{4}x\overline{E}_{p^{\prime }}(x)E_{p}(x)=(2\pi )^{4}\widehat{\delta }%
^{(4)}(p-p^{\prime })\equiv (2\pi )^{4}\delta _{kk^{\prime }}\delta
(p_{0}-p_{0}^{\prime })\delta (p_{2}-p_{2}^{\prime })\delta
(p_{3}-p_{3}^{\prime })  \label{e16}
\end{equation}
and complete

\begin{equation}
\sum\limits_{k}\int d^{3}pE_{p}(x)\overline{E}_{p}(y)=\sum\limits_{k}\int
dp_{0}dp_{2}dp_{3}E_{p}(x)\overline{E}_{p}(y)=(2\pi )^{4}\delta ^{(4)}(x-y)
\label{e17}
\end{equation}
Here we have used $\overline{E}_{p}(x)=\gamma ^{0}E_{p}^{\dagger }\gamma
^{0}.$

Using Eqs. (\ref{e8}) and (\ref{e9}) in (\ref{e7}), the last one can be
expressed as

\[
\Gamma _{f}^{(1)}=-i\int d^{4}xd^{4}y\sum\limits_{k}\int \frac{d^{3}p}{%
\left( 2\pi \right) ^{4}}\sum\limits_{k^{\prime }}\int \frac{d^{3}p^{\prime }%
}{\left( 2\pi \right) ^{4}}Tr\{E_{p}\left( x\right) \left( \gamma .\overline{%
p}+m_{0}\right) \overline{E}_{p}\left( y\right) 
\]
\begin{equation}
\times E_{p^{\prime }}\left( y\right) \left( \frac{1}{\gamma .\overline{p}%
^{\prime }+\Sigma (\overline{p}^{\prime })}\right) \overline{E}_{p^{\prime
}}\left( x\right) \}  \label{e18}
\end{equation}

Making use of the property (\ref{e16}) one can easily integrate in $y$ and $%
p^{\prime }$ to obtain

\begin{equation}
\Gamma _{f}^{(1)}=-i\int d^{4}x\sum\limits_{k}\int \frac{d^{3}p}{\left( 2\pi
\right) ^{4}}Tr\left\{ E_{p}\left( x\right) \left( \frac{\gamma .\overline{p}%
+m_{0}}{\gamma .\overline{p}+\Sigma (\overline{p})}\right) \overline{E}%
_{p}\left( x\right) \right\}  \label{e19}
\end{equation}

At this point we need to consider the structure of the mass operator $\Sigma
\ $introduced in ref. \cite{vv2}

\begin{equation}
\mathop{\textstyle \sum }%
(\overline{p})=Z_{_{\Vert }}(\overline{p})\gamma \cdot \overline{p}_{_{\Vert
}}+Z_{\bot }(\overline{p})\gamma \cdot \overline{p}_{_{\bot }}+m(\overline{p}%
)  \label{e20}
\end{equation}
where $\overline{p}_{_{\Vert }}=(p_{0,}0,0,p_{3})$, $\overline{p}_{_{\bot
}}=(0,0,-\sqrt{2gBk},0).$ The coefficients $Z_{_{\Vert }}(\overline{p}),$ $%
Z_{\bot }(\overline{p})$and $m(\overline{p})$ are functions of $\overline{p}%
^{2}.$ Notice the usual separation in the presence of a magnetic field
between parallel and perpendicular variables. Then, taking into account (\ref
{e20}), the contribution (\ref{e19}) can be written as

\begin{equation}
\Gamma _{f}^{(1)}=-i\int d^{4}x\sum\limits_{k}\int \frac{d^{3}p}{\left( 2\pi
\right) ^{4}}Tr\left\{ E_{p}\left( x\right) \left( \frac{\gamma .\overline{p}%
+m_{0}}{(1+Z_{_{\Vert }})\gamma .\overline{p}_{\Vert }+(1+Z_{\bot })\gamma
\cdot \overline{p}_{_{\bot }}+m(\overline{p})}\right) \overline{E}_{p}\left(
x\right) \right\}  \label{e21}
\end{equation}

The integral in $x$ yields

\begin{equation}
\Gamma _{f}^{(1)}=-i\left( 2\pi \right) ^{4}\delta
^{(3)}(0)\sum\limits_{k}\int \frac{d^{3}p}{\left( 2\pi \right) ^{4}}%
Tr\left\{ \frac{\gamma .\overline{p}+m_{0}}{(1+Z_{_{\Vert }})\gamma .%
\overline{p}+(1+Z_{\bot })\gamma \cdot \overline{p}_{_{\bot }}+m(\overline{p}%
)}\right\}  \label{e22}
\end{equation}
where the notation $\delta ^{(3)}(k)=\delta (k_{0})\delta (k_{2})\delta
(k_{3})$ is understood. After taking the trace, integrating in $p_{2}$ and
doing the Wick rotation to Euclidean coordinates, we obtain

\begin{equation}
\Gamma _{f}^{(1)}=8\pi gB\delta ^{(4)}(0)\sum\limits_{k}\int dp_{4}dp_{3}%
\frac{(1+Z_{_{\Vert }})\overline{p}_{\Vert }^{2}+(1+Z_{\bot })\overline{p}%
_{_{\bot }}^{2}+m(\overline{p})m_{0}}{(1+Z_{_{\Vert }})^{2}\overline{p}%
_{\Vert }^{2}+(1+Z_{\bot })^{2}\overline{p}_{_{\bot }}^{2}+m^{2}(\overline{p}%
)}  \label{e25}
\end{equation}

We are interested in the contribution of $\Gamma _{f}^{(1)}$ to the minimum
equations (\ref{e3}) and (\ref{e4}). In the case of the gap equation, such a
contribution can be found directly from Eq. (\ref{e7}), differentiating with
respect to $\overline{G}.$ On the other hand, the contribution of $\Gamma
_{f}^{(1)}$ to the scalar vev extremum equation takes the form

\begin{equation}
\frac{\partial \Gamma _{f}^{(1)}}{\partial \varphi _{c}}=8\pi \lambda
_{y}gB\delta ^{(4)}(0)\sum\limits_{k}\int dp_{4}dp_{3}\frac{m(\overline{p})}{%
\overline{p}_{\Vert }^{2}+(1+Z_{\bot })^{2}\overline{p}_{_{\bot }}^{2}+m^{2}(%
\overline{p})}  \label{e25a}
\end{equation}
where the solution\footnote{%
The demostration that $Z_{_{_{\Vert }}}=0$ is a solution of the gap equation
in the present theory can be done along the same line of reasoning followed
in the Appendix of the first paper of ref.\cite{vv1}.} $Z_{_{_{\Vert }}}=0$
of the SD equation (\ref{e3}) was explicitly used.

At large magnetic field, the main contribution to the sum in $k$ comes from
the LLL, i.e. $\overline{p}_{_{\bot }}^{2}=2gBk=0$. Then, in this
approximation Eq. (\ref{e25a}) becomes

\begin{equation}
\frac{\partial \Gamma _{f}^{(1)}}{\partial \varphi _{c}}=8\pi \lambda
_{y}gB\delta ^{(4)}(0)\int dp_{4}dp_{3}\frac{m(\overline{p}_{\Vert })}{%
\overline{p}_{\Vert }^{2}+m^{2}(\overline{p}_{\Vert })}  \label{e25b}
\end{equation}

In general, the dynamical mass $m(\overline{p}_{\Vert })$ depends on the
momentum. However, we can expect that, similarly to QED \cite
{mirans-gus-shoko},\cite{composite}, $m(\overline{p}_{\Vert })$ behaves as a
constant in the infrared region, and diminishes with increasing $\left| 
\overline{p}_{_{_{_{\Vert }}}}\right| $. Therefore, the main contribution to
the integral in Eq. (\ref{e25b}) will come from the infrared region $\left| 
\overline{p}_{_{_{_{\Vert }}}}\right| <\sqrt{gB}.$ From the above
discussion, it is reasonable to approximate the function $m(\overline{p}%
_{\Vert })$ by a constant solution $m(\overline{p}_{\Vert })\approx m(o)=m$
and use $\sqrt{gB}$ as a natural cutoff. This leads us to the final result

\begin{equation}
\frac{\partial \Gamma _{f}^{(1)}}{\partial \varphi _{c}}=V^{(4)}\frac{%
\lambda _{y}}{2\pi ^{2}}gBm\ln \left( \frac{gB}{m^{2}}\right)
=-V^{(4)}\lambda _{y}<\overline{\psi }\psi >  \label{e26a}
\end{equation}
where $V^{(4)}$ represents an infinite four dimensional volume, and $<%
\overline{\psi }\psi >=iTr\left\{ \overline{G}(x,x)\right\} =-\frac{gBm}{%
2\pi ^{2}}\ln \left( \frac{gB}{m^{2}}\right) $ denotes the
fermion-antifermion condensate \cite{ackley},\cite{smilga} induced by the
external magnetic field.

It is worth to notice the following. It is a well known fact that in the
present model, in the absence of a magnetic field, the effective action
(potential) has a non-zero minimum at the one-loop level, but this minimum
lies far outside the expected range of validity of the one-loop
approximation, even for arbitrarily small coupling constant, so it must be
rejected as an artifact of the used approximation \cite{coleman-weinb}. When
a magnetic field is present, the term $\frac{\partial \Gamma _{f}^{(1)}}{%
\partial \varphi _{c}},$ being proportional to the fermion condensate,
dominates the radiative corrections in the minimum equation (\ref{e4}). Due
to this, the dynamically generated fermion condensate gives rise to a non
zero scalar minimum that is in agreement with the used approximation. In
other words, thanks to the magnetic field, a consistent minimum solution can
be generated by radiative corrections. In this sense, a sort of
non-perturbative Coleman-Weinberg mechanism takes place, with the difference
that in the present case no dimensional transmutation is needed. Since the
theory already contains a dimensional parameter, the magnetic field $B$,
there is no need to include scalar-gauge interactions in order to trade a
dimensionless coupling for the dimensional parameter $\varphi _{c}$ \cite
{coleman-weinb}. No constraint between the couplings has to be assumed,
except that they are all sufficiently weak.

The two-loop contributions are a little more involved. As we have not enough
space in a letter to give all the detailed calculations, we will explicitly
show, for the sake of understanding, the evaluation of one term. The others
can be found in a similar way. The complete calculation will be published
elsewhere.

First, notice that the second and fourth term in Eq. (\ref{e6}) generate
tadpole diagrams in the SD equation (\ref{e3}). It is easy to realize that
the tadpole diagram with the gauge-fermion vertex vanishes. However, the
tadpole associated to the scalar-fermion vertex is not zero when $\varphi
_{c}\neq 0$ and, as shown below, it has a significant contribution to the
gap equation. Let us evaluate this tadpole contribution, which we denote by $%
\sum\nolimits^{T}.$ 
\begin{equation}
\mathop{\textstyle \sum }%
\nolimits^{T}(x,y)=i\frac{\delta \Gamma _{2}^{^{T}}}{\delta G}=-i\lambda
_{y}^{2}\delta ^{4}(x-y)\int d^{4}z\Delta (x-z)tr\left[ \overline{G(}%
z,z)\right]  \label{26}
\end{equation}

We can transform Eq. (\ref{26}) to momentum space with the help of the $%
E_{p}\left( x\right) $ functions to obtain

\[
\int d^{4}xd^{4}y\overline{E}_{p}\left( x\right) 
\mathop{\textstyle \sum }%
\nolimits^{T}(x,y)E_{p^{\prime }}\left( y\right) =(2\pi )^{4}\widehat{\delta 
}^{(4)}(p-p^{\prime })%
\mathop{\textstyle \sum }%
\nolimits^{T}(\overline{p}) 
\]
\begin{eqnarray}
&=&-i\lambda _{y}^{2}\int d^{4}xd^{4}z\overline{E}_{p}\left( x\right) \int 
\frac{d^{4}q}{\left( 2\pi \right) ^{4}}\frac{e^{iq\left( x-z\right) }}{%
q^{2}+M^{2}+i\in }  \nonumber \\
&&\times \sum\limits_{k"}\int \frac{d^{3}p^{"}}{\left( 2\pi \right) ^{4}}%
Tr\left\{ E_{p"}\left( z\right) \left( \frac{1}{\gamma .\overline{p}"+\sum (%
\overline{p}")}\right) \overline{E}_{p"}\left( z\right) \right\}
E_{p^{\prime }}\left( x\right)  \label{e27}
\end{eqnarray}

Taking into account that \cite{ackley}

\begin{eqnarray}
\int d^{4}xe^{iqx}\overline{E}_{p}\left( x\right) E_{p^{\prime }}\left(
x\right) &=&\left( 2\pi \right) ^{4}\delta ^{\left( 3\right) }(p^{\prime
}+q-p)e^{iq_{1}(p_{2}^{\prime }+p_{2})/2gB}e^{-\widehat{q}_{\bot }^{2}/2} 
\nonumber \\
&&\times \sum\limits_{\sigma \sigma ^{\prime }}\frac{e^{i(n-n^{\prime
})\varphi }}{\sqrt{n(k,\sigma )!n^{\prime }(k^{\prime },\sigma ^{\prime })!}}%
J_{nn^{\prime }}(\widehat{q}_{\bot })\Delta (\sigma )\delta _{\sigma \sigma
^{\prime }}  \label{e28}
\end{eqnarray}
one can integrate in $x$ and $z$ to find

\[
(2\pi )^{4}\widehat{\delta }^{(4)}(p-p^{\prime })\sum\nolimits^{T}(\overline{%
p})=-i\lambda _{y}^{2}\int d^{4}q\sum\limits_{k"}\int d^{3}p^{"}\delta
^{\left( 3\right) }(q)\delta ^{\left( 3\right) }(p^{\prime }+q-p) 
\]
\[
\times e^{-\widehat{q}_{\bot }^{2}}\frac{e^{iq_{1}(p_{2}^{\prime
}+p_{2}-2p_{2}")/2gB}}{q^{2}+M^{2}+i\in }\sum\limits_{\sigma }\frac{%
e^{i(n-n^{\prime })\varphi }}{\sqrt{n(k,\sigma )!n^{\prime }(k^{\prime
},\sigma )!}}J_{nn^{\prime }}(\widehat{q}_{\bot })\Delta (\sigma ) 
\]
\begin{equation}
\times \sum\limits_{\sigma "}\left\{ Tr\left[ \Delta \left( \sigma "\right) 
\frac{1}{\gamma .\overline{p}"+\sum (\overline{p}")}\right] \frac{1}{%
n"(k",\sigma ")!}J_{n"n"}(\widehat{q}_{\bot })\right\} ,  \label{e29}
\end{equation}
This equation can be further simplified after taking the trace and using the
small $\widehat{q}_{\bot }^{2}$approximation of the $J-$functions 
\begin{equation}
J_{nn"}(\widehat{q}_{\bot })\rightarrow \frac{\left[ \max (n,n")\right] !}{%
\left| n-n"\right| !}\left[ i\widehat{q}_{\bot }\right] ^{\left| n-n"\right|
}\rightarrow n!\delta _{nn"},  \label{e30}
\end{equation}
which can be justified by the presence of the exponential factor $e^{-%
\widehat{q}_{\bot }^{2}}$ in the integrand of Eq. (\ref{e29}). Moreover,
thanks to the delta $\delta ^{\left( 3\right) }(q)$, the integrations in $%
q_{0},$ $q_{2},$ $q_{3}$ are trivial. Thus, from the previous considerations
and using the properties of the $\Delta $ matrices \cite{ackley}, we arrive
at 
\[
(2\pi )^{4}\widehat{\delta }^{(4)}(p-p^{\prime })\sum\nolimits^{T}(\overline{%
p})=-2i\lambda _{y}^{2}\delta ^{\left( 4\right) }(p^{\prime }-p)\int
dq_{1}\sum\limits_{k"}\int d^{3}p^{"}e^{-\widehat{q}_{\bot }^{2}}\frac{%
e^{iq_{1}(p_{2}^{\prime }+p_{2}-2p_{2}")/2gB}}{q_{1}^{2}+M^{2}+i\in } 
\]
\begin{equation}
\times \left\{ \frac{2m(\overline{p}")}{\left( 1+Z_{_{_{\Vert }}}\right) ^{2}%
\overline{p}_{_{_{_{\Vert }}}}"^{2}+\left( 1+Z_{\bot }\right) ^{2}\overline{p%
}_{_{\bot }}"^{2}+m^{2}(\overline{p}")}\right\}  \label{e31}
\end{equation}

Finally, after integrating in $q_{1}$ and $p_{2}",$ and transforming to
Euclidean space, we get 
\begin{equation}
\mathop{\textstyle \sum }%
\nolimits^{T}=\frac{\lambda _{y}^{2}}{2\pi ^{3}}\frac{gB}{M^{2}}%
\sum\limits_{k"}\int dp_{4}^{"}dp_{3}^{"}\frac{m(\overline{p}")}{\left(
1+Z_{_{_{\Vert }}}\right) ^{2}\overline{p}_{_{_{_{\Vert }}}}"^{2}+\left(
1+Z_{\bot }\right) ^{2}\overline{p}_{_{\bot }}"^{2}+m^{2}(\overline{p}")}
\label{32}
\end{equation}

Similarly to the analysis done in the one-loop case, it can be seen that the
main contribution to Eq. (\ref{32}) comes from the $k"=0$ term of the sum.
Taking into account this and using the solution $Z_{_{_{\Vert }}}=0,$ the
tadpole contribution to the gap equation reduces to

\begin{equation}
\mathop{\textstyle \sum }%
\nolimits^{T}=\frac{\lambda _{y}^{2}}{2\pi ^{3}}\frac{gB}{M^{2}}\int
dp_{4}^{"}dp_{3}^{"}\frac{m(\overline{p}_{_{_{\Vert }}}")}{\overline{p}%
_{_{_{_{\Vert }}}}"^{2}+m^{2}(\overline{p}_{_{_{\Vert }}}")}  \label{32a}
\end{equation}

One can recognize here the same integral that led us to the fermion
condensate in Eq. (\ref{e25b}). Therefore, we find 
\begin{equation}
\mathop{\textstyle \sum }%
\nolimits^{T}\simeq \frac{1}{\pi ^{2}}\frac{\lambda _{y}^{2}}{\lambda
\varphi _{c}^{2}}gBm\ln \left( \frac{gB}{m^{2}}\right) =-2\frac{\lambda
_{y}^{2}}{\lambda \varphi _{c}^{2}}<\overline{\psi }\psi >,  \label{e33}
\end{equation}

Note that the tadpole term is proportional to the magnetic field and
inversely proportional to the scalar mass. This functional dependence will
be responsible for the notorious increment of the mass solution in this
model as compared to other theories, as discussed below.

Taking into account the leading contributions to Eqs. $\left( \ref{e3}%
\right) $ and $\left( \ref{e4}\right) $ at l$\arg $e magnetic field, one
arrives at the following minimum equations for the fermion mass and the
scalar vev respectively, 
\begin{equation}
m\simeq m_{0}+\left( \frac{g^{2}}{4\pi }-\frac{\lambda _{y}^{2}}{8\pi }%
\right) \frac{m}{4\pi }\ln ^{2}\left( \frac{gB}{m^{2}}\right) +\frac{1}{\pi
^{2}}\frac{\lambda _{y}^{2}}{\lambda \varphi _{c}^{2}}gBm\ln \left( \frac{gB%
}{m^{2}}\right)   \label{34}
\end{equation}
\begin{equation}
\frac{\lambda }{6}\varphi _{c}^{3}+\frac{\lambda ^{2}}{64\pi ^{2}}\varphi
_{c}^{3}\left( \ln \left( \frac{\varphi _{c}^{2}}{gB}\right) -\frac{11}{3}%
\right) -\lambda _{y}\frac{gB}{2\pi ^{2}}m\ln \left( \frac{gB}{m^{2}}\right)
\simeq 0  \label{35}
\end{equation}

Eq. $\left( \ref{35}\right) $ can be further simplified by noting that at
large $B$ one can neglect the terms $\sim \lambda ^{2}$ coming from the
one-loop scalar self-interaction, compared to the term coming from the
fermion condensate contribution $\sim $ $<\overline{\psi }\psi >$. Then, the
scalar minimum satisfies

\begin{equation}
\varphi _{c}^{3}\simeq \frac{\lambda _{y}}{\lambda }\frac{3gB}{\pi ^{2}}m\ln
\left( \frac{gB}{m^{2}}\right)  \label{36}
\end{equation}

Similarly, the gauge and scalar bubble diagram contributions to the gap
equation, (second term in $\left( \ref{34}\right) $), are negligible
compared to the tadpole contribution, (third term in $\left( \ref{34}\right) 
$), so the equation becomes

\begin{equation}
m\simeq m_{0}+\frac{1}{\pi ^{2}}\frac{\lambda _{y}^{2}}{\lambda \varphi
_{c}^{2}}gBm\ln \left( \frac{gB}{m^{2}}\right)  \label{36a}
\end{equation}

Substituting $\left( \ref{36}\right) $ in Eq. $\left( \ref{36a}\right) ,$ it
is found 
\begin{equation}
m\simeq \frac{1}{\sqrt{\kappa }}\sqrt{gB}  \label{37}
\end{equation}
where the coefficient $\kappa $ satisfies

\begin{equation}
\kappa \ln \kappa \simeq 1.4\frac{\lambda }{\lambda _{y}^{4}}  \label{38}
\end{equation}

The corresponding solution for the Higgs vev is 
\begin{equation}
\varphi _{c}\approx \frac{0.8}{\kappa ^{1/2}\lambda _{y}}\sqrt{gB}
\label{e39}
\end{equation}

It is easy to check that for weak couplings the solutions $\left( \ref{37}%
\right) $ and $\left( \ref{e39}\right) $ are in agreement with the used
approximations. Note that there is no zero solution for the scalar vev in
this large field approximation. Both, the minimum of the scalar field and
the dynamically generated mass are driven by the external magnetic field. As
the dynamical fermion mass breaks the discrete chiral symmetry $\left( \ref
{e2}\right) ,$ this model might be considered as one more example of the
phenomenon of magnetic catalysis. However, if the current model is extended
to include complex scalars (complex scalars do not change at all the
conclusions of this paper) and non-simple gauge groups, as $SU(2)\times U(1)$
in the electroweak model, the symmetry breaking phenomenon will have a
different nature. There, we do not have chiral symmetry, but the
magnetic-field-driven scalar minimum will break the gauge symmetry by giving
mass to the gauge fields coupled to it. Thus, as we claimed at the beginning
of the paper, in richer models with scalar fields, the magnetic field can
either reinforce gauge symmetry breaking if it is already broken, or induce
it, through non-perturbative effects.

Comparing the induced fermion dynamical mass Eq. $\left( \ref{37}\right) $
with the mass generated when no scalar field is present \cite
{mirans-gus-shoko},\cite{ackley}, $m\simeq \sqrt{\left| gB\right| }\exp
\left[ -2\pi /g\right] ,$ or with the one obtained when the couplings are
fine-tuned so the scalar vev is set to zero \cite{vv1}, $m\approx \sqrt{%
\left| gB\right| }\exp \left[ -2\pi \sqrt{1/\left( g^{2}+\frac{\lambda
_{y}^{2}}{2}\right) }\right] ,$ one realizes that the scalar interactions,
when taken into account in a self-consistent way, can dramatically
strengthen the generation of mass. This observation is easy to corroborate
by direct evaluation of the mass $\left( \ref{37}\right) $ for typical
values of Yukawa coupling $\lambda _{y}$ and scalar self-coupling $\lambda $%
. For instance, if we take $\lambda _{y}=0.7$, which is the approximate
value of the Yukawa coupling for the top quark, and $\lambda =0.4,$ which
corresponds to a Higgs mass of $115\ Gev$, we find $m\simeq 0.6\sqrt{\left|
gB\right| }.$ The same Yukawa coupling, on the other hand, gives just $%
m\simeq 10^{-5}\sqrt{\left| gB\right| }$ if the scalar vev is fine-tuned to
zero\cite{vv1}. In the case of pure gauge-fermion interactions, as for
instance in QED, the generated mass would be even much smaller \cite
{mirans-gus-shoko},\cite{ackley}.

A non-simple group extension of the model discussed in this work could be of
interest as an effective theory in condensed matter problems, where $%
SU(2)\times U(1)$ gauge theories (although without Higgs fields) have been
previously proposed to describe the rich phase structure of high T$_{c}$
superconductors \cite{mav}.

It seems, however, that the most immediate physical extension of the present
model would be the electroweak theory. This case is particularly interesting
in the light of the recent works on the role of magnetic fields in
electroweak baryogenesis \cite{Giovannini-Shaposhnikov}-\cite{Skalazub}. In
view of the magnetic-field-driven non-perturbative enlargement of the Higgs
vev, Eq. $\left( \ref{e39}\right) $, it is possible that in the electroweak
theory the increase in the scalar vev will be large enough to guarantee the
baryogenesis condition $\left( \ref{1}\right) .$ It remains therefore as an
open question whether the effect found in this paper can influence the
recent conclusions \cite{Giovannini-Shaposhnikov}-\cite{Skalazub} about
baryogenesis in the presence of primordial magnetic fields.

\begin{description}
\item  
\begin{center}
{\bf ACKNOWLEDGMENTS }
\end{center}
\end{description}

It is a pleasure to thank V. Gusynin and V. Miransky for enlightening
discussions on the phenomenon of magnetic catalysis, and M. Laine, M.
Shaposhnikov and G. Veneziano for useful discussions and remarks. The
authors wish to acknowledge the warm hospitality extended to them at the
University of Barcelona (VI), and the Institute for Space Studies of
Catalonia (EF), during the last stage of this work. This work has been
supported in part by NSF grant PHY-9722059 (EF and VI) and NSF POWRE grant
PHY-9973708 (VI).

\end{document}